\documentclass{article}
\usepackage{ijcai21}

\usepackage{makecell}
\usepackage{times}
\usepackage{soul}
\usepackage{url}
\usepackage[hidelinks]{hyperref}
\usepackage[utf8]{inputenc}
\usepackage[small]{caption}
\usepackage{graphicx}
\usepackage{amsmath}
\usepackage{amsthm}
\usepackage{booktabs}
\usepackage{algorithm}
\usepackage{latexsym}
\usepackage{soul}
\usepackage[utf8]{inputenc}
\usepackage{amsmath}
\usepackage{booktabs}
\usepackage{amsthm}
\usepackage{thmtools}
\usepackage{thm-restate}

\usepackage[utf8]{inputenc} % allow utf-8 input
\usepackage[T1]{fontenc}    % use 8-bit T1 fonts
\usepackage{url}            % simple URL typesetting
\usepackage{booktabs}       % professional-quality tables
\usepackage{amsfonts}       % blackboard math symbols
\usepackage{nicefrac}       % compact symbols for 1/2, etc.
\usepackage{microtype}      % microtypography
\usepackage{color}

\def\vec#1{\mathchoice{\mbox{\boldmath$\displaystyle#1$}}
  {\mbox{\boldmath$\textstyle#1$}}
  {\mbox{\boldmath$\scriptstyle#1$}}
  {\mbox{\boldmath$\scriptscriptstyle#1$}}}
\newcommand{\mat}[1]{\mathbf{#1}}

\usepackage{helvet}  %Required
\usepackage{courier}  %Required
\usepackage{subfig}
\usepackage{enumitem}
\usepackage{multirow}
\usepackage{mathtools}
\usepackage{balance}
\usepackage{flushend}

\graphicspath{{figures/}}

\DeclareMathOperator*{\argmax}{argmax}
\usepackage{amssymb}
\usepackage{tabularx}
\usepackage{hhline}
\usepackage{algpseudocode}
\usepackage{array}
\usepackage{listings}
\usepackage{diagbox}
\usepackage{tikz}
\usepackage[utf8]{inputenc}
\usepackage{bm}
\usetikzlibrary{spy}

\title{A Novel Attribute Reconstruction Attack in Federated Learning}

\author{
Lingjuan Lyu$\textsuperscript{\rm 1}$
\and
Chen Chen$\textsuperscript{\rm 2}$
\affiliations
$\textsuperscript{\rm 1}$ Sony AI
$\textsuperscript{\rm 2}$ Zhejiang University
\emails
lingjuanlvsmile@gmail.com, cc33@zju.edu.cn
}

\begin{document}

\maketitle
\begin{abstract}
Federated learning (FL) emerged as a promising learning paradigm to enable a multitude of participants to construct a joint ML model without exposing their private training data. Existing FL designs have been shown to exhibit vulnerabilities which can be exploited by adversaries both within and outside of the system to compromise data privacy. However, most current works conduct attacks by leveraging gradients on a small batch of data, which is less practical in FL. In this work, we consider a more practical and interesting scenario in which participants share their epoch-averaged gradients (share gradients after at least 1 epoch of local training) rather than per-example or small batch-averaged gradients as in previous works. We perform the first systematic evaluation of attribute reconstruction attack (ARA) launched by the malicious server in the FL system, and empirically demonstrate that the shared epoch-averaged local model gradients can reveal sensitive attributes of local training data of any victim participant. To achieve this goal, we develop a more effective and efficient gradient matching based method called cos-matching to reconstruct the training data attributes. We evaluate our attacks on a variety of real-world datasets, scenarios, assumptions. Our experiments show that our proposed method achieves better attribute attack performance than most existing baselines. 
\end{abstract}

\section{Introduction}
Federated learning (FL) allows multiple participants to collaboratively construct a better global model~\cite{mcmahan2017communication}. In FL, each participant trains a local model on its own data and periodically shares model parameters or updates with a parameter server. Since training data never leave the participants, FL provides a privacy-aware solution for scenarios where data is sensitive (e.g., biomedical records, private images, personal text and speech, and personally identifiable information like location, purchase etc.). However, recent works have pointed out that FL may not provide sufficient privacy guarantees~\cite{kairouz2019advances,lyu2020threats,lyu2020privacy}, as communicating model updates throughout the training process can nonetheless leak private training data information, from shallow~\cite{melis2019exploiting,hitaj2017deep} to deep leakage~\cite{zhu2019deep,zhao2020idlg}. 

For shallow leakage from gradients, ~\cite{hitaj2017deep} used GANs~\cite{goodfellow2014generative} to synthesize images that look similar to the training data from the gradient. However, this attack only works when all class members look alike (e.g., face recognition).~\cite{melis2019exploiting} developed learning-based methods to demonstrate that these exchanged updates can leak unintended information about participants' training data. It should be noted that all these attacks require true class labels.

For deep leakage from gradients,~\cite{zhu2019deep} demonstrated \emph{Deep Leakage from Gradients} (DLG) and proposed an optimization algorithm that can obtain both the training inputs and the labels in few iterations. A follow-up work called \emph{Improved Deep Leakage from Gradients} (iDLG)~\cite{zhao2020idlg} pointed out that DLG has difficulty in convergence and discovering the ground-truth labels consistently. However, there are several inherent weaknesses in both DLG and iDLG that may limit their applicabilities in FL, including: (1) both works adopted a second-order optimization method called L-BFGS, which is more computationally expensive compared with the first-order optimization we adopted; (2) both works are only applicable to the gradient computed on a small batch of samples, i.e., at most B=8 in DLG, and B=1 in iDLG, which are less practical in FL, in which gradient is normally shared after at least 1 epoch of local training on each participant's local data for communication efficiency~\cite{mcmahan2017communication}; (3) both works used untrained model (pre1 in our work), neglecting gradients over multiple communication rounds.

In addition to the above risks in existing literatures, in this work, we further ask the question: is it possible to steal other private information from epoch-averaged gradients instead of small batch-averaged gradients? To answer this question, we initiate a new attack called \emph{attribute reconstruction attack} (ARA), i.e., we aim to accurately reconstruct the sensitive attributes of the training examples of the victim participant by exploiting the privacy vulnerabilities of the released local model gradients.

In summary, the following main contributions are made: 

 \begin{itemize}
 \item We initiate a more practical and interesting attack against FL system, called \emph{attribute reconstruction attack} (ARA), in which participants share their gradients/updates after at least 1 epoch of local training instead of small batches. 

 \item We propose a more effective and efficient gradient matching based method called cos-matching, which utilizes the cosine similarity between the true update from the victim participant and the virtual update. 

 \item We conduct a comprehensive analysis and investigate various practical settings, including both the isolated and non-isolated settings, and the adversary with a range of prior knowledge (attribute distribution, membership and true label). Extensive experiments validate the effectiveness of our method.
 \end{itemize}

\section{Problem Definition}
\begin{figure}
     \centering
     \includegraphics[width=0.45\textwidth]{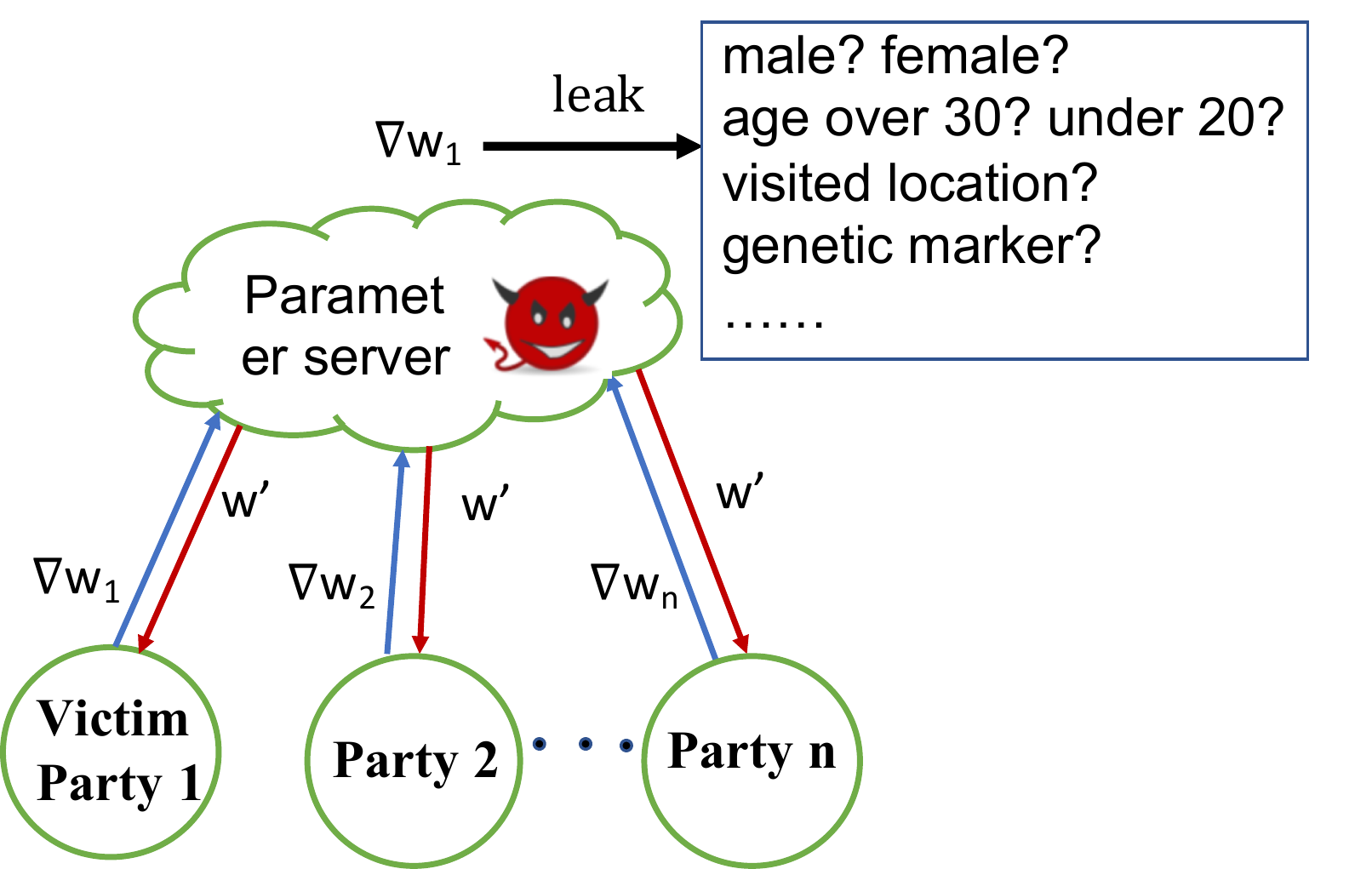}
     \caption{Attribute reconstruction attack (ARA) in federated learning. The malicious server can reconstruct victim participant's attributes from its local model updates.}
     \label{fig:ARA_pipeline}
     \vspace{-5mm}
 \end{figure}

Attribute inference or reconstruction attack aims to determine the value of a particular attribute given that the adversary knows all the other non-sensitive attributes of one record and has access to either a trained model~\cite{melis2019exploiting} or model embedding~\cite{song2020information}. However, none of the previous works systematically investigate attribute reconstruction attack (ARA) in FL. To fill in this gap, we establish attribute reconstruction attack in FL, which targets on the sensitive attributes of the training samples in any victim participant's data set, as illustrated in Fig~\ref{fig:ARA_pipeline}.

The intuition behind ARA is that the observations of local model gradients can be used to infer sensitive attributes, which are in turn based on the participants' private training data. The snapshots of the gradients across communication rounds can also be exploited to generate different views of the target participant's local training data, thus inferring the \emph{exact} value of the sensitive attributes. We remark that our objective of reconstructing the exact value of $x_s$ is different from model inversion attack~\cite{yeom2018privacy}, in which the attacker aims to learn the private attribute distribution $p(x_s|y, x_{ns})$ given all the non-sensitive attributes $x_{ns}$ and the true label $y$. Our attack assumption is more practical and reasonable in FL setting. We did not assume that the adversary has access to the joint distribution of the private feature and label $p(X_s, X_{ns}, Y)$ as in~\cite{yeom2018privacy}. More importantly, we suppose \textbf{each participant will send local model updates after 1 epoch of local training on its whole data}, rather than updates on a small batch of data as in previous works~\cite{zhu2019deep,zhao2020idlg}. Moreover, the attacker may not know the membership of the target record as a prior knowledge.

 \section{Attack Methodologies}
 \label{sec:attack_method}
 In FL, both the server and participants may be malicious, and compromise the FL system. In particular, we focus on the malicious server who aims to infer the sensitive attributes (usually discrete)~\cite{song2020information}, which are normally binary or categorical demographic related attributes. This malicious server can use the target participant's model updates revealed in each round of the collaborative training process and adaptively update the reconstructed attribute.

Concretely, each record of interest can have $K$ candidate records by enumerating the unknown sensitive attribute value from the corresponding attribute value range of size $K$.
 For any unknown attribute $x_s$ within a limited range (categorical or discretizing each numeric attribute into bins of equal width), the attacker can enumerate the target attribute $x_s$ with all the possible values (with size $K$) for each record $\vec{x}$, resulting in $K$ candidate records/guess points.

 \subsection{Attacker types and prior knowledge}
In this work, we consider both the active attacker and the passive attacker.
 The active attacker can isolate the target participant, and create a local view of the model for it. In this scenario, the attacker can isolate the target participant and segregate the target participant's learning process~\cite{nasr2019comprehensive}.
 By contrast, the passive attacker strictly follows the FL protocol, and sends the aggregated parameters of all participants to the victim participant, but tries to reconstruct the sensitive attributes of the victim participant's training records.

In real scenarios, the attacker may have no prior knowledge about the membership before conducting ARA. We next explore several practical scenarios based on the attacker's prior knowledge about the membership.

 %%%%%%%%%%%%%%%%%%%%%%%%%%%%%%%%%%%%%
 \begin{algorithm}[t]
 \caption{Attribute reconstruction based on cos-matching}\label{Algorithm:m_dummy}
 \begin{algorithmic}[1]
 \State \textbf{Input:} 
 Victim participant records $(\mat{X},\vec{Y})$ of size $(N,d+1)$ and $(N,1)$. The $i$-th column of $\mat{X}=\{X[:,:i-1],X_s,X[:,i+1:]\}$ is unknown (corresponding to the sensitive attribute $X_s$); Each element of $X_s$ falls into $\{1,\cdots,K\}$; Victim participant target model $f(w_t)$ and loss function $\ell$; Gradient $\nabla w_t$ of $f$ computed w.r.t. $\mat{X}$; Gumbel softmax parameter $\gamma$; attribute prior.
 \State \textbf{Output:} Predicted sensitive attribute $X_s$.
 \State \textbf{Execute:} 
 \If{prior==known}
 \State $X_s \leftarrow log(prior)$
 \Else 
 \State $X_s \leftarrow \mathcal{N}(0,1)$ 
 \EndIf
 \State Generate multinomial distribution for the sensitive attribute  $X_s$ in $\mat{X}$, by computing the gumbel softmax 
 of $X_s$, i.e., Softmax($X_s/\gamma$,dim=1).
 \State Compute the predicted attribute value $a=\sum_{n=1}^K n[Softmax(X_s/\gamma,dim=1)]_n$.
 \State Concatenate the predicted attribute with other non-sensitive attributes to get virtual input $X'=Concat(X[:,:i-1],a,X[:,i+1:])$.
 \State Given the true gradients for $T$ epochs, the attacker solves the following optimization problem w.r.t. $X_s$ (refer to Equation~\ref{equ:w_t}, Equation~\ref{equ:X_s}):
 \[
 X_s^*=\mathrm{\argmax_{X_s}} \sum_{t=1}^T \mathrm{cosinesim}(\frac{\partial \ell(X',Y; w_t)}{\partial w_t},\nabla {w_t})
 \]
 \State \Return predicted attribute $X_s$
 \end{algorithmic}
 \end{algorithm}
 %%%%%%%%%%%%%%%%%%%%%%%%%%%%%%%%%%%%%
 
 \subsection{When membership is known}  
 \label{sec:known_membership}
 Given membership information (the true label $Y$ is known), the attacker can directly reconstruct any attribute by using our cos-matching method as shown in Algorithm~\ref{Algorithm:m_dummy}. Here cosine similarity is used to measure the angular distance between the true update $\nabla w_t$ from the victim participant and the virtual update $\nabla {w_t'}$. The objective function is formulated as:
 \begin{equation}
 J=cosinesim(\nabla {w_t'},\nabla w_t)
 \end{equation}

 We denote the local data of the victim participant as: $\mat{X} = [X_{ns},X_s]$, where $X_{ns}=[X[:,:i-1],X[:,i+1:]]$ represents the non-sensitive attributes, while $X_s=X[:,i]$ represents the sensitive/target attribute at $i$-th index.
 Based on the prior knowledge of the attribute of interest, we consider two cases to initialize the sensitive attribute: 1) if the prior distribution is known, then we initialize the virtual attribute from the prior distribution $p$ of the target attribute over $K$ potential values; 2) if the prior distribution is unknown, then we initialize the virtual attribute from a random distribution $\mathcal{N}(0,1)$. We then generate the multinomial distribution for the sensitive attribute, and compute the predicted attribute value as follows:
 \begin{equation}
 \label{equ:multinomial}
 a=\sum_{n=1}^K n[Softmax(X_s/\gamma,dim=1)]_n
 \end{equation}

 We adopt Gumbel-softmax~\cite{jang2016categorical} to compute the multinomial distribution parameter over all the potential attribute values (line 9 Algorithm~\ref{Algorithm:m_dummy}), where $\gamma$ is the Gumbel softmax parameter.
 Afterwards, we concatenate the predicted attribute with other non-sensitive attributes to get “virtual inputs” $X'=Concat(X[:,:i-1],a,X[:,i+1:])$, then feed these “virtual inputs” into the victim participant's model and get the “virtual gradients”.
 \begin{equation}
 \label{equ:w_t}
 \nabla {w_t'} = \frac{\partial \ell(X',Y; w_t)}{\partial w_t}
 \end{equation}

 Matching the virtual gradients with the received gradients from the victim participant in order to update the target attribute can be cast as an optimization problem. When the virtual gradients become closer to the original gradients, virtual attribute also becomes closer to the real training data attribute. Given gradients at a certain iteration $t$, we obtain the training data attribute by maximizing the following objective:
 \begin{equation}
 \label{equ:X_s}
 X_s^*=\mathrm{\argmax_{X_s} cosinesim}(\nabla {w_t'},\nabla w_t)
 \end{equation}

The objective function \emph{cosinesim} is differentiable w.r.t the target attribute $X_s$, and thus can be optimized using the standard gradient-based methods. Note that our proposed cos-matching method is also applicable to the scenario when the true label $Y$ is unknown, the only modification to Algorithm~\ref{Algorithm:m_dummy} is to initialize the virtual label from either the label prior or a random distribution. 

\subsection{When membership is unknown}
 \label{sec:member_unknown}
 If the membership information is unknown, we first need to determine the membership of the target data points by launching a membership inference attack (MIA). To achieve this goal, we propose to apply a \emph{Gaussian Mixture Model} (GMM) with two components (Member and Non-member) on the last-layer gradient variances of all the candidate records (generated by enumerating the unknown sensitive attribute from the attribute value range) to distinguish members from non-members. Our design rationale include: (1) we found that the gradient distribution for the member and non-member instances are distinguishable. In particular, the gradient variances of all the candidate training examples (Member) are all prone to 0, while the gradient variances of all the candidate test examples (Non-member) generally follow a more uniform distribution; (2) the gradients from the last layer leak the most membership information, as the last layer already contains the membership information that leaks from the output of the previous layers~\cite{nasr2019comprehensive}.

 After we derive the membership information from GMM, the problem becomes similar as the scenario when the membership is known (Section~\ref{sec:known_membership}). 
 
 \section{Experimental Setup}

 \subsection{Datasets}

 \textbf{Purchase.} 
 This dataset contains 600 different products (attributes), and each user has a binary record which indicates whether she has bought each of the products (a total of 197,324 data records). The records are clustered into 100 classes based on the similarity of the purchases, and our objective is to identify the class of each user's purchases~\cite{shokri2017membership}. For Purchase100 dataset, we randomly choose two attributes with different distributions, which correspond to 130-th and 595-th attribute. We further generate a Purchase2 dataset with 2 classes to investigate how ARA performs when the true label is unknown (Section~\ref{sec:label_unknown}).

 \textbf{Genome.} We implement a public genome dataset called HS3D (Homo Sapiens Splice Sites Dataset)\footnote{\url{http://www.sci.unisannio.it/docenti/rampone/}} for splice site prediction. It is a binary classification task in computational genetics which determines whether the target nucleotide sequence contains certain functional unit. The prepared dataset consists of total 28800 negative and 2880 positive samples. All the genome sequences are of length 20 ($\#$attributes=20). Similarly, for Genome dataset, we randomly choose two attributes with different attribute distributions, which correspond to 7-th and 17-th attribute. 

 \textbf{Location.} We use a curated location dataset from the publicly available set of mobile users' location “check-ins” in the Foursquare social network~\cite{shokri2017membership}. Each record in the resulting dataset has 446 binary attributes, representing whether the user visited a certain region or location type. We cluster the location dataset into 30 classes, each representing a different geosocial type. The classification task is to predict the user's geosocial type given his or her record.

 For all datasets, we randomly select 10\% records as public set, and partition the rest data into training set and test set with a ratio of 8:2. 
 A summary of all datasets is given in Table~\ref{tbl:datasets_summary}.

 \begin{table}[t]
 \scalebox{0.8}{
 \footnotesize
     \centering
     \begin{tabular}{l|cccccc}
     \toprule
     Dataset & \multicolumn{2}{c}{Purchase100} & \multicolumn{2}{c}{Location} & \multicolumn{2}{c}{Genome} \\
     \midrule
     Target $y$ & \multicolumn{2}{c}{\makecell{purchase \\type}} & \multicolumn{2}{c}{\makecell{geosocial \\type}}  & \multicolumn{2}{c}{\makecell{sequence \\detection}} \\
     Attribute $X_s$ & 130-th & 595-th & 1-st & 69-th & 7-th & 17-th \\ 
     $X_s$ variance & 26.20e+8 & 11.85e+8 & 2.6e+6 & 2070 & 138019 & 77610 \\ 
     Cramer's V & 0.2302 & 0.2310 & 0.1747 & 0.0412 & 0.0017 & 0.0016 \\ \bottomrule
     \end{tabular}
 }
 \caption{\footnotesize Summary of datasets. 
     Cramer's V captures statistical correlation between $Y$ and $X_s$ (0 indicates no correlation and 1 indicates perfectly correlated).}
     \label{tbl:datasets_summary}
 \end{table}

 \subsection{Experiment Configuration} 
 \label{sec:setup}
 For all datasets, we take \emph{Multi-Layer Perceptron} (MLP) model. For each participant in FL, we
 train an MLP model with one hidden layer of 128 units. We take ReLU (rectifier linear units) as the activation function, and SGD as the optimizer. We vary batch size from $\{8,32,|D_v|\}$, where $|D_v|$ refers to the local data size of the victim participant. For practicality, we adopt FedAvg protocol~\cite{mcmahan2017communication} and suppose each participant sends local updates after 1 epoch of local training. We set the maximum training epochs to 100. For cos-matching optimization, we adopt Adam optimizer. For all datasets, we set Gumbel softmax parameter via grid search.

In FL, each participant only has a limited amount of local data, henceforth, according to the available data size, we vary the local data size of each participant from \{50, 100, 500, 1000\} for Purchase100 and Genome datasets, and \{50, 100, 300\} for Location dataset. We consider gradients of different epochs chosen from $\{1, \cdots, 100\}$: untrained model (pre1); first 2 epochs (pre2); first 5 epochs (pre5); 5 chosen epochs from \{10, 20, 30, 40, 50\} (gap10); last 5 epochs (last5).

 \textbf{Evaluation Metrics}. For ARA, we use \emph{reconstruction accuracy}, which is defined as the fraction of the correct reconstruction for member and non-member attribute. When the membership is unknown, we compute \emph{MIA attack accuracy} as the fraction of the correct membership predictions.

 \subsection{Baselines}
 \label{sec:baseline}
 \textbf{Random guess (random)}: The attacker randomly picks a value from all the possible values of size $K$ as the predicted attribute, i.e., reconstruction accuracy=1/$K$.

 \textbf{L2-BFGS~\cite{zhu2019deep}}: The sensitive attribute is optimized to minimize the L2 distance between virtual gradients and real gradients. 

 \textbf{Attack model trained on public set}: If the attacker has access to some auxiliary data with true attribute values, the attacker can train a multi-class attack model which takes the non-sensitive attributes $\{(x_1, \cdots, x_d)\}$ of the public data as inputs, and uses the true attribute value $\{x_s\}$ as labels. After the attack model is trained, the attacker can predict the sensitive attribute of any record of interest. 

 \textbf{Inference from statistics (stats)}: When the true label is known, the server can directly infer the true attribute from the statistics calculated on all the enumerated records of interest without performing any optimization. In statistics based method, the attacker can store checkpoints of the victim participant's model across all epochs (suppose 1 epoch corresponds to 1 communication round). For all guess points, the attacker uses the local model of the victim participant to produce a statistic vector of size $K$ in each epoch. In this way, the attacker derives a statistic matrix of size $K*E$, with each row corresponding to a guess point associated with one possible attribute value, and each column corresponding to one epoch. 

 After these statistic matrices are built, the attacker can make decisions on all guess points based on the following heuristics: (1) label status matrix: choose row with the max number of correct predictions across epochs; (2) probability score matrix: choose row with the max probability sum across all epochs; (3) loss norm matrix: choose row with the min loss norm sum across all epochs; (4) final loss matrix: choose row with the min final loss; (5) last-layer gradient norm (gradient w.r.t. $w$ on each guess point $x$): choose row with the max gradient norm sum across all epochs; (6) gradient true label matrix: last-layer gradient connecting to the true label which corresponds to the only negative gradient value. Choose the guess point with the largest gradient value (closer to 0).
(7) majority: choose the majority row from above statistics.

 We remark that statistics based baselines require the prior knowledge of the true label and need to calculate per-record gradient using victim participant's model, while our reconstruction method has no such constraints.

 \section{Experimental Results}
 For experimentation, we consider different capabilities of the attacker, including membership=\{known, unknown\}, attribute prior=\{known, unknown\}, and true label=\{known, unknown\}. We also investigate three scenarios: \{victim participant isolate=1; victim participant isolate=0, $\#$participant=2; victim participant isolate=0, $\#$participant=10\}. Moreover, we study the efficacy of the iterative communication rounds in FL by integrating the gradients of different states \{pre1, pre2, pre5, gap10, last5\} (Sec~\ref{sec:setup}). We further investigate different local data size $|D_v|=\{50, 100, 300, 500, 1000\}$ and different local training batch size B=$\{8,32,|D_v|\}$.

 \subsection{When membership is known}
 \label{sec:Results_memberknown}
 For all datasets, we first consider the following scenario for illustration purpose: isolate=1, membership=known, true label=known, and local training batch size B=$|D_v|$. Table~\ref{tbl:Purchase_baseline_130th}, Table~\ref{tbl:Location_baseline_1st_69th} and Table~\ref{tbl:Genome_baseline_7th}
 list the attribute reconstruction accuracy for Purchase100, Location, and Genome datasets respectively. In particular, we list the best result for the stats based method and our method across \{pre1, pre2, pre5, gap10, last5\} epochs (Sec~\ref{sec:setup}). 

 As evidenced by all tables, using our cos-matching method to predict the attribute mostly outperforms the baselines of \{random, stats\}. Moreover, our method outperforms the baseline models trained on 100/1000 public examples (\{public100, public1000\}) in most cases. This implies that relying on the correlations of only a handful of labeled public examples to train an attack model is not sufficient to reconstruct the extract attribute value. More importantly, our proposed cos-matching method performs better than L2-BFGS for most datasets in most scenarios. We hypothesise that the reason is that Euclidean distance may not be effective in evaluating the difference between high dimensional vectors due to the curse of dimensionality. Moreover, Euclidean distance fails to capture the directional difference between gradient vectors. By contrast, cosine similarity is well-known to be more robust in evaluating similarity between high dimensional model parameters than Euclidean distance~\cite{huang2021personalized}.
 However, we also notice that L2-BFGS using L2 distance and L-BFGS performs better than our method on Genome, when local data size $|D_v|=\{50,100\}$, as shown in Table~\ref{tbl:Genome_baseline_7th}. We speculate that this is due to the fact that L2-BFGS is more suitable to the dataset with smaller number of attributes, but wider range. For example, Genome dataset only has 20 attributes, and the range length of the chosen 7-th and 17-th attributes equals 4, rather than 2 as in Purchase100 and Location.  
 %%%%%%%%%%%%%%%%%%%%%%%%%%%%%%%%%%%%%%
 \begin{table}[t]
 \caption{130-th and 595-th attribute reconstruction accuracy for Purchase100 when isolate=1, membership=known, prior=known, and true label=known. Best results are shown in bold.}
 \label{tbl:Purchase_baseline_130th}
 \centering
 \scalebox{0.65}{
 \begin{tabular}{|l|l|l|l|l|l|l|l|l|}
 \hline
 Attribute &\multicolumn{4}{|l|}{130-th (var=26.20e+8)} &\multicolumn{4}{|l|}{595-th (var=11.85e+8)} \\
 \hline
 $|D_v|$ &50 &100 &500 & 1000 &50 &100 &500 & 1000 \\ 
 \hline
 random  &0.5 &0.5 &0.5 &0.5  &0.5 &0.5 &0.5 &0.5 \\ 
 \hline
 public100 &0.8600 &0.8400 &0.8160 &0.7880 &0.6000 &0.7000 &0.6940 &0.7090 \\
 \hline
 public1000 &0.8200 &0.8500 &0.8260 &0.8240 &0.7000 &0.7600 &0.7340 &0.7090\\ 
 \hline 
 stats &0.6600 &0.7400 &0.6500 &0.6560 &0.7400 &0.7800 &0.6800 &0.6560\\ 
 \hline 
 L2-BFGS &1.0000 &0.9600 &0.7740 &0.7650 &1.0000 &\textbf{1.0000} &0.7240 &0.7330\\ 
 \hline
 Our & \textbf{1.0000} & \textbf{1.0000} & \textbf{0.9240}  & \textbf{0.8310} & \textbf{1.0000} & 0.9900 & \textbf{0.8080} & \textbf{0.7590}\\ 
 \hline
 \end{tabular}
 }
 \end{table}

 %%%%%%%%%%%%%%%%%%%%%%%%%%%%%%%%%%%%%%

 \begin{table}[t]
 \caption{1-st and 69-th attribute reconstruction accuracy for Location when isolate=1, membership=known, prior=known, and true label=known. Best results are shown in bold.}
 \label{tbl:Location_baseline_1st_69th}
 \centering
 \scalebox{0.72}{
 \begin{tabular}{|l|l|l|l|l|l|l|}
 \hline
 Attribute &\multicolumn{3}{|l|}{1-st (var=2.6e+6)} &\multicolumn{3}{|l|}{69-th (var=2070)} \\
 \hline
 $|D_v|$ &50 &100 & 300 &50 &100 & 300 \\ 
 \hline
 random  &0.5 &0.5 &0.5 &0.5  &0.5 &0.5  \\ 
 \hline
 public100 &0.8600 &0.8500 &0.8200 
 &0.5800 &0.5700 &0.5600 \\
 \hline
 public1000 &0.8800 &0.8400 &0.8367 
 &0.6000 &0.5800 &0.6433 \\
 \hline 
 stats &0.7800 &0.7600 &0.6900 &0.7800 &0.7000 &0.6567 \\ 
 \hline 
 L2-BFGS &1.0000 &1.0000 &0.8500 &1.0000 &0.7900 &0.7100 \\ 
 \hline
 Our & \textbf{1.0000} & \textbf{1.0000} & \textbf{0.8633}
 & \textbf{1.0000} & \textbf{0.9500} & \textbf{0.7967} \\ 
 \hline
 \end{tabular}
 }
 \end{table}
 %%%%%%%%%%%%%%%%%%%%%%%%%%%%%%%%%%%%%%
 \begin{table}[t]
 \caption{7-th and 17-th attribute reconstruction accuracy for Genome when isolate=1, membership=known, prior=known, and true label=known. Best results are shown in bold.}
 \label{tbl:Genome_baseline_7th}
 \centering
 \scalebox{0.65}{
 \begin{tabular}{|l|l|l|l|l|l|l|l|l|}
 \hline
 Attribute &\multicolumn{4}{|l|}{7-th (var=138019)} &\multicolumn{4}{|l|}{17-th (var=77610)} \\
 \hline
 $|D_v|$ &50 &100 &500 & 1000 &50 &100 &500 & 1000 \\ 
 \hline
 random  &0.25 &0.25 &0.25 &0.25  &0.25 &0.25  &0.25 &0.25    \\ 
 \hline
 public100 &0.2600 &0.3300 &0.2500 &0.2620 &0.2800 &0.2500 &0.2680 &0.2660 \\
 \hline
 public1000 &0.3200 &0.3100 &0.2760 &0.2840 &0.3400 &0.3300 &0.2840 &0.2830 \\ 
 \hline 
 stats &0.2400 &0.2800 &0.3420 &0.3210 &0.2800 &0.3000 &0.3120 &0.2950 \\ 
 \hline 
 L2-BFGS &\textbf{1.0000} &\textbf{0.9100} &0.3180 &0.2700 &\textbf{1.0000} &\textbf{0.7300} &0.2820 &0.2690 \\ 
 \hline
 Our & 0.5600 & 0.5200 & \textbf{0.3540} & \textbf{0.3550} & 0.5800 & 0.4800 & \textbf{0.3720} & \textbf{0.3350} \\ 
 \hline
 \end{tabular}
 }
 \end{table}
 %%%%%%%%%%%%%%%%%%%%%%%%%%%%%%%%%%%%%%
 %%%%%%%%%%%%%%%%%%%%%%%%%%%%%%%%%%%%%%
 \textbf{Impact of the attribute distribution:} 
 We find that attribute distribution is closely related to the attribute reconstruction accuracy.
 Compared with attribute with higher variance, attribute with lower variance is harder to attack. For instance, from Table~\ref{tbl:Purchase_baseline_130th} for Purchase100 dataset, attribute reconstruction accuracy on the 130-th attribute with higher variance is higher than the 595-th attribute. From Table~\ref{tbl:Location_baseline_1st_69th} for Location dataset, attribute reconstruction accuracy on the 1-st attribute with higher variance is higher than the 69-th attribute. From Table~\ref{tbl:Genome_baseline_7th} for Genome dataset, 7-th attribute with slightly higher variance is also slightly higher than the 17-th attribute, especially when the local data size is larger.

 \textbf{Impact of the participant isolation:} As shown in Table~\ref{tbl:ARA_isolation}, compared with isolate=1, isolate=0 generally delivers slightly lower reconstruction accuracy. We articulate that the reason is: in isolate=0, the server does not isolate the victim participant, hence the downloaded global model in each communication round integrates all participants' model updates, weakening attack efficacy.

 \begin{table}[t]
 \caption{Impact of participant isolation on ARA when membership=known, prior=known, and true label=known (We fix local batch size B=$|D_v|$=100).}
 \label{tbl:ARA_isolation}
 \centering
 \scalebox{0.68}{
 \begin{tabular}{|c|c|c|c|c|c|c|}
 \hline
  &\multicolumn{6}{c|}{ARA} \\ 
  \hline
 Data &\multicolumn{2}{c|}{Purchase100} & \multicolumn{2}{c|}{Genome} &\multicolumn{2}{c|}{Location} 
 \\ 
 \hline
 Attribute  &130 &595 &7 &17  &1 &69 
 \\
 \hline
 isolate=1  &\textbf{1.0000} &\textbf{0.9900} &\textbf{0.5200} &\textbf{0.4800} &\textbf{1.0000}  &\textbf{1.0000} 
 \\ 
 \hline
 isolate=0, participant=2  &1.0000 &0.8900 &0.4200 &0.3700 &1.0000  &1.0000 
 \\ 
 \hline
 isolate=0, participant=10  &0.9700 &0.8600 
 &0.4200 &0.3700 &1.0000  &1.0000 
 \\ 
 \hline
 \end{tabular}
 }
 \end{table}

 \textbf{Impact of the local batch size:} There is a consistent conclusion on the relationship between local model batch size and attack performance. As evidenced by Table~\ref{tbl:ARA_batch}, for the investigated batch size B=$\{8,32,|D_v|\}$, B=$|D_v|$ mostly outperforms B=$\{8,32\}$, and the attack performance generally becomes better with larger batch size. We attribute this to the more randomness introduced by the smaller batch size during local model training, the released local model gradients cannot capture the full pass over the whole local data, making it harder to conduct ARA.

 \begin{table}[t]
 \caption{Impact of batch size on ARA when isolate=1, membership=known, prior=known, and true label=known (We fix the local data size of the victim participant as $|D_v|=100$, and vary the victim participant local training batch size B=$\{8,32,|D_v|\}$).}
 \label{tbl:ARA_batch}
 \centering
 \scalebox{0.7}{
 \begin{tabular}{|c|c|c|c|c|c|c|c|}
 \hline
  &\multicolumn{6}{c|}{ARA} \\ 
  \hline
 Data &\multicolumn{2}{c|}{Purchase100} & \multicolumn{2}{c|}{Genome} &\multicolumn{2}{c|}{Location} 
 \\ 
 \hline
 Attribute  &130 &595 &7 &17  &1 &69 \\
 \hline
 B=8  &0.8500 &0.6900 &\textbf{0.4000} &0.3500 &0.9800  &0.9000 
 \\ 
 \hline
 B=32  &0.9600 &0.8000 &0.3900 &0.4300 &0.9300  &0.7700 
 \\ 
 \hline
 B=$|D_v|$  &\textbf{1.0000} &\textbf{0.9900} &0.3300 &\textbf{0.4600} &\textbf{1.0000}  &\textbf{0.9500} 
 \\ 
 \hline
 \end{tabular}
 }
 \end{table}

 \textbf{Impact of the training phases:} Attack performance differs a lot when the gradients of different epochs are considered. Generally, the gradients of the previous epochs (\{pre1, pre2, pre5\}) outperform the gradients of later epochs (\{gap10, last5\}), as manifested in Table~\ref{tbl:ARA_phases}. Our interpretation is that at the beginning of the training process, the gradients are large such that more useful information can be utilized for ARA. With the evolution of training, local model tends to converge thus gradients decrease, leading to weak signals contained in the released gradients.

 \begin{table}[t]
 \caption{Impact of training phases on ARA when isolate=1, membership=known, prior=known, and true label=known (We fix local batch size B=$|D_v|$=100).}
 \label{tbl:ARA_phases}
 \centering
 \scalebox{0.85}{
 \begin{tabular}{|c|c|c|c|c|c|c|}
 \hline
  &\multicolumn{6}{c|}{ARA} \\ 
  \hline
 Data &\multicolumn{2}{c|}{Purchase100} & \multicolumn{2}{c|}{Genome} &\multicolumn{2}{c|}{Location} 
 \\ 
 \hline
 Attribute  &130 &595 &7 &17  &1 &69 
 \\
 \hline
 pre1  &1.0000 &0.9500 &0.3300 &0.4500 &1.0000  &0.9200 
 \\ 
 \hline
 pre2  &1.0000 &0.9700 &\textbf{0.3300} &\textbf{0.4600} &1.0000  &0.9400 
 \\ 
 \hline
 pre5  &\textbf{1.0000} &0.9700 &0.3200 &0.4200 &\textbf{1.0000}  &\textbf{0.9500} 
 \\ 
 \hline
 gap10  &0.9800 &\textbf{0.9900} &0.2600 &0.3000 &0.9800  &0.8100 
 \\ 
 \hline
 last5  &0.9300 &0.9300 &0.1900 &0.3200 &0.9600  &0.8100 
 \\ 
 \hline
 \end{tabular}
 }
 \end{table}

 \textbf{Impact of the local data size:} Local data size also impacts attack performance. Generally, ARA becomes less effective when the local data size increases, as demonstrated in Table~\ref{tbl:ARA_datasize}. This is because the released gradients contain the information for all the examples, while attribute reconstruction is inherently a post-hoc per-example task. Therefore, it becomes harder to reconstruct per-example attribute when the victim participant has more data.

 \begin{table}[t]
 \caption{Impact of local data size $|D_v|$ when isolate=1, membership=known, prior=known, and true label=known (We adopt gradients of pre5, and fix the victim participant batch size B=$|D_v|$).}
 \label{tbl:ARA_datasize}
 \centering
  \scalebox{0.82}{
 \begin{tabular}{|c|c|c|c|c|c|c|c|}
 \hline
  &\multicolumn{6}{c|}{ARA} \\ 
  \hline
 Data &\multicolumn{2}{c|}{Purchase100} & \multicolumn{2}{c|}{Genome} &\multicolumn{2}{c|}{Location} \\
 \hline
 Attribute  &130 &595 &7 &17 &1 &69 \\
 \hline
 $|D_v|=50$  &\textbf{1.0000} &\textbf{1.0000} &\textbf{0.4400} &\textbf{0.4800} &\textbf{1.0000} &\textbf{1.0000}\\ 
 \hline
 $|D_v|=100$  &1.0000 &0.9700 &0.3200 &0.4200 &1.0000 &0.9400\\ 
 \hline
 $|D_v|=500$  &0.9000 &0.7840 &0.2940 &0.2960 &- &-\\ 
 \hline
 $|D_v|=1000$  &0.8310 &0.7480 &0.2750 &0.2550 &- &-\\ 
 \hline
 $|D_v|=300$  &- &- &- &- &0.8667 &0.7967\\ 
 \hline
 \end{tabular}
 }
 \end{table}

 \textbf{Impact of model complexity:}
 To investigate how model complexity impacts ARA, we use Location data and three fully connected models, with hidden layer sizes of 128, 1024, \{1024, 256\} respectively. 
 As shown in Table~\ref{tbl:ARA_Location_model_complexity}, the model with higher capacity is more vulnerable to ARA. This partly reflects the fact that the model with higher capacity can memorize more information about the training data~\cite{zhang2017understanding}. 

 %%%%%%%%%%%%%%%%%%%%%%%%%%%%%%%%%%%%%%

 \begin{table}[!htbp]
 \caption{Impact of model complexity. ARA for Location when isolate=1, membership=known, prior=known, and true label=known (We fix the victim participant local training batch size B=$|D_v|$). Best results are shown in bold.}
 \label{tbl:ARA_Location_model_complexity}
 \centering
 \scalebox{0.84}{
 \begin{tabular}{|l|l|l|l|l|l|l|}
 \hline
 Attribute &\multicolumn{3}{|l|}{1-st (var=2.6e+6)} &\multicolumn{3}{|l|}{69-th (var=2070)} \\
 \hline
 $|D_v|$ &50 &100 & 300 &50 &100 & 300 \\ 
 \hline
 128  & 1.0000 & 1.0000 & 0.8633& 1.0000 & 0.9500 & 0.7967  \\ 
 \hline
 1024 &1.0000 &1.0000 &1.0000 &1.0000 &1.0000 &0.9633 \\
 \hline
 \{1024,256\} &\textbf{1.0000} &\textbf{1.0000} &\textbf{1.0000} &\textbf{1.0000} &\textbf{1.0000} &\textbf{0.9833}  \\
 \hline
 \end{tabular}
 }
 \end{table}

 %%%%%%%%%%%%%%%%%%%%%%%%%%%%%%%%%%%%%

 \begin{table}[htp]
 \caption{Purchase100 MIA using GMM when isolate=1, prior=known, true label=known, but membership=unknown (For the victim participant, we vary its local data size from $|D_v|$=\{50, 100, 500, 1000\}, and local batch size B=$\{8,32,|D_v|\}$).}
 \label{tbl:MIA_Purchase}
 \centering
 \scalebox{0.65}{
 \begin{tabular}{|c|c|c|c|c|c|c|c|c|c|}
 \hline
  &\multicolumn{8}{c|}{MIA} \\ 
  \hline
 $|D_v|$ &\multicolumn{2}{c|}{50} &\multicolumn{2}{c|}{100} & \multicolumn{2}{c|}{500} & \multicolumn{2}{c|}{1000} \\ 
 \hline
 Attribute  &130 &595 &130 &595  &130 &595 &130 &595 \\
 \hline
 B=8  &0.9900 &0.9900 &0.9925  &0.9800 &0.8860  &0.8865 &0.8245  &0.8205  \\ 
 \hline
 B=32  &0.9700  &\textbf{0.9950} &0.9900  &0.9950 &0.9005  &0.9110 &0.8160 &0.7990   \\ 
 \hline
 B=$|D_v|$  & \textbf{0.9900} & 0.9900 & \textbf{1.0000} & \textbf{1.0000} & \textbf{0.9805} & \textbf{0.9800} & \textbf{0.9568} & \textbf{0.9568} \\ 
 \hline
 \end{tabular}
 }
 \end{table}

 %%%%%%%%%%%%%%%%%%%%%%%%%%%%%%%%%%%%%

 \begin{table}[htp]
 \caption{Location MIA using GMM when isolate=1, prior=known, true label=known, but membership=unknown (For the victim participant, we vary local data size of the victim participant from $|D_v|$=\{50, 100, 300\}, and local batch size B=$\{8,32,|D_v|\}$).}
 \label{tbl:MIA_Location}
 \centering
 \scalebox{0.85}{
 \begin{tabular}{|c|c|c|c|c|c|c|c|}
 \hline
  &\multicolumn{6}{c|}{MIA} \\ 
  \hline
 $|D_v|$ &\multicolumn{2}{c|}{50} &\multicolumn{2}{c|}{100} & \multicolumn{2}{c|}{300} \\ 
 \hline
 Attribute  &1 &69 &1 &69  &1 &69 \\
 \hline
 B=8  &0.9650 &\textbf{0.9850} &0.9450  &\textbf{0.9650} &0.7958  &0.8000   \\ 
 \hline
 B=32  &0.9850  &0.9800 &0.9325  &0.9375 &0.8000  &0.8000  \\ 
 \hline
 B=$|D_v|$  & \textbf{0.9850} & 0.9800 & \textbf{0.9600} & 0.9575 & \textbf{0.8958} & \textbf{0.8933} \\ 
 \hline
 \end{tabular}
 }
 \end{table}
 %%%%%%%%%%%%%%%%%%%%%%%%%%%%%%%%%%%%%
 \subsection{When membership is unknown}
 When the membership information is unknown, we first conduct membership inference attack (MIA) by following our proposed GMM based method in Section~\ref{sec:member_unknown}. In particular, we focus on Purchase100 and Location datasets under the scenario of isolate=1 for illustration purpose. As shown in Table~\ref{tbl:MIA_Purchase}, on Purchase100 dataset, based on the last-layer gradient variances calculated on all the candidate records with the enumerated 130-th attribute, we get MIA accuracy of $\{0.9900,1.0000,0.9805,0.9568\}$ for \{50,100,500,1000\} randomly sampled member and non-member records respectively. Similarly, for 595-th attribute, we get MIA accuracy of $\{0.9900,1.0000,0.9800,0.9568\}$ for \{50,100,500,1000\} randomly sampled member and non-member records respectively. These results significantly outperform the accuracy of 73.4\% reported in~\cite{nasr2019comprehensive}, which relies on a more complex Encoder-Decoder attack model. On Location dataset with enumerated 1-st or 130-th attribute, we also get relatively high MIA accuracy, as shown in Table~\ref{tbl:MIA_Location}.
 % %%%%%%%%%%%%%%%%%%%%%%%%%%%%%%%%%%%%%

 After we determine the membership information, we follow Algorithm~\ref{Algorithm:m_dummy} to reconstruct the member attribute. We observe that the conclusions for the scenario when the membership is known (Section~\ref{sec:Results_memberknown}) also generally hold in the scenario when the membership is unknown.

 \subsection{When the prior distribution is unknown}
 We further investigate ARA when the prior distribution is unknown. In this scenario, we initialize the virtual attribute values from a random distribution $\mathcal{N}(0,1)$.
 Table~\ref{tbl:attacks_Location_dummy_prior0_isolate1_1st_69th} shows the 1-st and 69-th attribute reconstruction accuracy for Location dataset across different local training batch size B=$\{8,32,|D_v|\}$, when isolate=1, membership=known, true label=known, but prior=unknown. Although the attack performance is slightly lower than the attack performance when prior=known as in Table~\ref{tbl:Location_baseline_1st_69th}, the accuracy results are still mostly higher than the baselines of \{random, public100, public1000, stats\}. Similar observations also hold for Purchase100, by comparing Table~\ref{tbl:Purchase_baseline_prior0_130th} with Table~\ref{tbl:Purchase_baseline_130th}. 
 
 \begin{table}[!htbp]
 \caption{1-st and 69-th attribute reconstruction accuracy for Location when isolate=1, membership=known, true label=known, but prior=unknown. Best results are shown in bold.}
 \label{tbl:attacks_Location_dummy_prior0_isolate1_1st_69th}
 \centering
 \scalebox{0.85}{
 \begin{tabular}{|l|l|l|l|l|l|l|l|l|l|l|l|l|}
 \hline
 Attribute &\multicolumn{3}{|l|}{1-st (var=2.6e+6)} &\multicolumn{3}{|l|}{69-th (var=2070)} \\
 \hline
 $|D_v|$ &50 &100 & 300 &50 &100 & 300 \\ 
 \hline
 B=8  &0.8600 &0.8500 &\textbf{0.7833}
 &0.8600 &0.7100 &0.5967 \\ 
 \hline
 B=32 &0.9400 &0.8000 &0.7433 
 &0.8600 &0.6400 &0.6300 \\
 \hline
 B=$|D_v|$ &\textbf{0.9400} &\textbf{0.8800} &0.7667 
 &\textbf{0.9400} &\textbf{0.7400} &\textbf{0.6500} \\
 \hline 
 \end{tabular}
 }
 \end{table}

 %%%%%%%%%%%%%%%%%%%%%%%%%%%%%%%%%%%%%%
 \begin{table}[t]
 \caption{130-th and 595-th attribute reconstruction accuracy for Purchase100 when isolate=1, membership=known, true label=known, but prior=unknown. Best results are shown in bold.}
 \label{tbl:Purchase_baseline_prior0_130th}
 \centering
 \scalebox{0.65}{
 \begin{tabular}{|l|l|l|l|l|l|l|l|l|}
 \hline
 Attribute &\multicolumn{4}{|l|}{130-th (var=26.20e+8)} &\multicolumn{4}{|l|}{595-th (var=11.85e+8)} \\
 \hline
 $|D_v|$ &50 &100 &500 & 1000 &50 &100 &500 & 1000 \\ 
 \hline
 B=8  &0.8600 &0.7800 &0.6460 &0.5830 &0.6800 &0.7100 &0.6280 &0.6380 \\ 
 \hline
 B=32 &0.9000 &0.7600 &0.6520 &0.6440 &0.8400 &0.7500 &0.6820 &0.6480 \\
 \hline
 B=$|D_v|$ &\textbf{0.9400} &\textbf{0.8400}  &\textbf{0.7660} &\textbf{0.7390} &\textbf{0.9200} &\textbf{0.8500} &\textbf{0.7300} &\textbf{0.6900}\\ 
 \hline 
 \end{tabular}
 }
 \end{table}

 %%%%%%%%%%%%%%%%%%%%%%%%%%%%%%%%%%%%%%

 \subsection{When the true label is unknown}
 \label{sec:label_unknown}
When the true label is unknown, we can initialize the virtual label either from a random distribution if the label distribution is unknown, or from the prior label distribution if the label distribution is known. In this scenario, the attacker needs to optimize both the virtual attribute and virtual label in  Algorithm~\ref{Algorithm:m_dummy}. 
For experimentation, we investigate Purchase2 and Genome, all with 2 classes. For both datasets, our method can reconstruct the true label with 100\% accuracy. In particular, given the prior label distribution, it takes $<$200 iterations for the extract label reconstruction, while it takes longer when the label distribution is unknown. 
    
When isolate=1, membership=known, prior=known, but true label=unknown, for Purchase2, we achieve ARA of $\{1.0000, 1.0000, 0.8360, 0.7570\}$ for 130-th attribute with $|D_v|=\{50, 100, 500, 1000\}$ respectively. Similarly, we achieve ARA of $\{1.0000, 0.9800, 0.7300, 0.7380\}$ for 595-th attribute. For Genome, ARA performance is similar to the scenario when the true label is known (Table~\ref{tbl:Genome_baseline_7th}). 

 %%%%%%%%%%%%%%%%%%%%%%%%%%%%%%%%%%%%%%

 \section{Conclusion}
 In this work, we initiate a more practical and interesting attack against FL system, called \emph{attribute reconstruction attack} (ARA), in which participants share their epoch-averaged gradients instead of small batch-averaged gradients. In particular, we consider a malicious parameter server in the FL system. To launch ARA, we propose a more effective and efficient method called cos-matching, which utilizes the cosine similarity to measure the angular distance between the true update from the victim participant and the virtual update. Our method exhibits significantly improved performance compared to most existing baselines in all settings we studied. We also conduct a systematic exploration on the factors that may affect ARA. Extensive experiments on real-world datasets and comprehensive analysis offer new insights for designing privacy-preserving FL methods and systems. This work has not investigated the scenario when the attacker is one of the participants in FL system, which would immediately be our future work. 
\bibliographystyle{named}
\bibliography{references}

\end{document}